\documentclass[aps,prd,nofootinbib,preprintnumbers,floatfix,notitlepage,longbibliography]{revtex4-2}

\usepackage{amsmath,amsthm,amssymb}
\usepackage{color}
\usepackage[colorlinks=true,linkcolor=blue,citecolor=red,urlcolor=magenta]{hyperref}

\allowdisplaybreaks[1]

\begin{document}

\title{Higher-curvature gravity in Weyl geometry:\\ No-ghost and no-tachyon conditions}

\author{Tomoya Tachinami}
\email[Email: ]{tachinami(a)nc-toyama.ac.jp}
\affiliation{National Institute of Technology, Toyama College, Toyama, 939-8630, Japan}

\begin{abstract}
Weyl geometry, in which the length scale can be chosen independently at each point of spacetime, provides a natural framework for extending general relativity through Weyl invariance, that is, local scale symmetry.
We construct a gravitational theory whose Lagrangian is an arbitrary function of the scalar curvature in Weyl geometry and establish the conditions under which the theory is free of ghosts and tachyons around maximally symmetric vacua.
For an arbitrary function, Weyl invariance requires a compensating scalar field in addition to the metric and the Weyl vector.
By linearizing the action with an auxiliary field, we show that the Weyl vector couples to the scalar sector only through a single combination of the scalar fields.
Consequently, the Weyl vector, shifted by the logarithmic gradient of this combination, becomes a Weyl-invariant massive vector field without gauge fixing.
In the Einstein frame, given that the graviton is ghost-free, the no-tachyon condition for the Weyl vector, together with the no-ghost condition for the scalaron, the scalar mode associated with the higher-curvature terms, reduces to the positivity of the kinetic coefficient of the compensating scalar.
When this coefficient vanishes, the scalaron does not propagate, leaving only the graviton and the massive vector.
As an example, we show that for a Lagrangian containing constant, linear, and quadratic curvature terms, both a massive vector and a massive scalaron are necessarily present throughout the stability region, whereas without the linear term the scalaron becomes massless.
\end{abstract}

\maketitle

\section{Introduction}
General relativity, proposed by Einstein in 1915, describes gravity as the geometry of curved spacetime and has successfully accounted for a wide range of gravitational phenomena.
However, on cosmological scales, there remain observational facts that are difficult to explain within the framework of general relativity alone, most notably the existence of the dark sector, that is, dark matter and dark energy.
To address these problems, attempts to extend Riemannian geometry itself, the very foundation of general relativity, have been discussed.

Weyl geometry originates in the unified theory of gravity and electromagnetism proposed by Weyl in 1918~\cite{Weyl:1918ib}.
In addition to the metric tensor, Weyl geometry introduces a vector field, the Weyl vector, that governs the length gauge at each point as an independent geometric object.
Weyl attempted to identify this vector field with the four-potential of the electromagnetic field.
However, his unified theory failed because the scale of clocks and rulers depended on the path along which they were transported, the so-called second clock effect.
For the historical development, we refer the reader to Ref.~\cite{Scholz:2017pfo}.

Weyl geometry allows the length gauge to be chosen independently at each point of spacetime, providing a natural framework for extending general relativity from the viewpoint of local scale symmetry.
Weyl geometry has been revived in recent years from a perspective borrowed from elementary particle physics.
The quadratic curvature terms of Weyl geometry are invariant under Weyl transformations in four dimensions, and together with the kinetic term of the Weyl vector, they constitute Weyl's original action.
Linearizing the quadratic terms by introducing an auxiliary scalar field, one finds that this symmetry is spontaneously broken by a Stueckelberg-type mechanism of geometric origin~\cite{Ghilencea:2018dqd,Ghilencea:2019jux}.
The scalar field is absorbed by the Weyl vector, which then becomes massive, and the Einstein--Proca action emerges in the broken phase with a positive cosmological constant, no ghosts, and a conserved number of degrees of freedom.
In this picture, the Planck scale is not a fundamental parameter of the action but arises as the scale of spontaneous breaking, set by the vacuum expectation value of the scalar field, which is driven to a constant dynamically in an expanding universe.
The same framework has also been applied to inflation~\cite{Ghilencea:2019rqj,Ferreira:2019zzx}.

Based on this reformulation, the resulting theory has been confronted with observations across a remarkably wide range of scales.
In static spherical symmetry, the vacuum field equations admit an exact solution that generalizes the Schwarzschild metric~\cite{Yang:2022icz}, and a solution of this class has been used to model galactic rotation curves without particle dark matter~\cite{Burikham:2023bil}.
Compact objects constructed for a variety of equations of state turn out to be more massive than their general relativistic counterparts~\cite{Haghani:2023nrm}.
The generalized Friedmann equations have been compared with cosmological observational data and the $\Lambda$CDM model~\cite{Harko:2024fnt}.
The gravitational spin Hall effect, that is, the polarization-dependent propagation of light, has also been studied in Weyl geometry~\cite{Oancea:2023ylb}.
Weyl geometric gravity is therefore no longer a purely formal construction but a theory with a growing phenomenology.

The interpretation of the Weyl vector as dark matter deserves particular emphasis among these motivations.
A massive vector behaves as cold dark matter once the Hubble rate falls below its mass, and no coupling to the Standard Model beyond gravity need be postulated for it to be produced~\cite{Graham:2015rva,Ema:2019yrd}.
These analyses assume that the vector carries a constant mass during and after inflation, as for a Stueckelberg mass.
Weyl geometry supplies such a vector without introducing any new field, because the Weyl vector belongs to the connection rather than the matter content, and it acquires its mass through a Stueckelberg mechanism of geometric origin.
In Weyl quadratic-curvature inflation, however, this mass depends on the inflaton, which induces a direct coupling between the inflaton and the vector.
Once this coupling and reheating history are considered, the Weyl vector can account for the observed relic abundance depending on the reheating temperature~\cite{Tang:2020ovf,Wang:2022ojc}.

As an extension of general relativity in Riemannian geometry, $F(R)$ gravity, in which the Lagrangian is an arbitrary function of the Ricci scalar $R$, has been developed as a model of inflation, dark energy, and dark matter~\cite{Starobinsky:1980te,Amendola:2006we,Cembranos:2008gj}, and its viability criteria are well established.
The first derivative of the function must be positive, $F'>0$, so that the effective gravitational coupling is positive and neither the graviton nor the additional scalar degree of freedom, the scalaron, becomes a ghost.
Here, a prime denotes differentiation with respect to the argument of the function.
In the vacuum, the scalaron must in addition have a positive mass squared around the maximally symmetric solution.
For the de Sitter solution, this requires the condition $0<RF''/F'<1$, whereas for the Minkowski solution it reduces to the positivity of the second derivative $F''>0$~\cite{Faraoni:2007yn,Sokolowski:2007rd,Sotiriou:2008rp,DeFelice:2010aj}.
When matter fields are included, the curvature is no longer fixed at its vacuum value, and these conditions must hold over the whole range of curvature realized in the cosmic history.
In particular, in the high-density regime, the positivity of the scalaron mass reduces to that of the second derivative $F''>0$, whose violation leads to the Dolgov--Kawasaki instability~\cite{Dolgov:2003px,Faraoni:2006sy}.

In contrast, in gravitational theories based on Weyl geometry, specific functional forms have been treated in numerous studies~\cite{Ghilencea:2018dqd,Ghilencea:2019jux,Ghilencea:2019rqj,Ferreira:2019zzx,Yang:2022icz,Burikham:2023bil,Haghani:2023nrm,Harko:2024fnt,Oancea:2023ylb,Wang:2022ojc}.
In these works, the aim was inflation or a specific astrophysical application.
In Refs.~\cite{Ghilencea:2018dqd,Ghilencea:2019jux,Ghilencea:2019rqj,Ferreira:2019zzx,Wang:2022ojc}, the Weyl gauge was fixed by setting the scalar field, or a combination of scalar fields, to a constant, whose value then sets the Planck scale, while in several astrophysical applications the gauge condition was imposed on the Weyl vector~\cite{Yang:2022icz,Burikham:2023bil,Haghani:2023nrm}.
Tang and Wu~\cite{Tang:2020ovf} treated an arbitrary function of the scalar curvature of Weyl geometry and kept the coefficient of the scalar kinetic term free.
After fixing the Einstein gauge, they required the scalar to be a normal field, that is, to have a kinetic term of the correct sign, and argued that both signs of the scalar kinetic term lead to consistent theories and viable inflation.
However, the sign of the mass term of the Weyl vector was not examined, and the stability of the vacuum was not analyzed for an arbitrary function.
In the other works cited above, freedom from ghosts and tachyons was established, if at all, only for the specific functional form under consideration rather than for an arbitrary function, and the coefficient of the scalar kinetic term was fixed rather than left free. 
A further common feature is the difficulty of coupling ordinary matter.
Weyl invariance enforces a trace condition on the energy-momentum tensor, so that conformally invariant ordinary matter is of radiation type unless an explicit dependence of the matter Lagrangian on the Weyl vector is postulated~\cite{Haghani:2023nrm,Harko:2024fnt}.
Since that choice is model-dependent, and the healthiness of the gravitational sector is a prerequisite for all the applications above, we work throughout in vacuum.

Therefore, this study constructs higher-curvature gravity based on Weyl geometry for an arbitrary function and derives its no-ghost and no-tachyon conditions.
We first construct the theory in the Jordan frame and linearize the action by introducing an auxiliary field.
Along the way, we show that the Weyl vector couples to the scalar sector only through a single combination of the scalar fields. The Weyl vector, shifted by the logarithmic gradient of this combination, becomes a Weyl-invariant massive vector without any gauge fixing, and no kinetic scalar-vector mixing survives.
We then perform the conformal transformation to the Einstein frame and derive the healthiness conditions.

The paper is organized as follows.
In Sec.~\ref{sec:WG}, we review the basics of Weyl geometry.
In Sec.~\ref{sec:HCG}, we construct the action of higher-curvature gravity based on Weyl geometry and derive the no-ghost and no-tachyon conditions.
In Sec.~\ref{sec:QCG}, we study quadratic curvature gravity as an example.
Sec.~\ref{sec:concl} presents our conclusions and an outlook.

Throughout the paper, we use the natural units $c=\hbar=1$ and denote the reduced Planck mass by $M_p=(8\pi G)^{-1/2}$.
Greek indices of tensors such as $\mu,\nu,\cdots$ are of space-time, while Latin ones such as $i,j,\cdots$ are spatial.
The Minkowski metric is $\eta_{\mu\nu}=\mathrm{diag}(-1,1,1,1)$.
The symbol $\partial_\mu$ denotes partial differentiation $\frac{\partial}{\partial x^\mu}$.
The Riemann tensor is defined as $R^\mu{}_{\nu\rho\sigma}=\partial_\rho\Gamma^\mu{}_{\nu\sigma}-\cdots$.
The function $f$ denotes the Weyl geometric Lagrangian function of $\mathcal R$, while $F$ denotes the Lagrangian function of the Ricci scalar $R$ in the Riemannian theory quoted for comparison.
A prime denotes differentiation with respect to the argument of the function.

In this study, by a Weyl transformation, we mean the operation that rescales the metric and, at the same time, transforms every field according to its Weyl charge, while transforming the Weyl gauge field inhomogeneously.
This is a local gauge symmetry, and the action considered here is invariant under it.
By a conformal transformation, on the other hand, we mean a field redefinition that rewrites only the metric, leaving all other fields untouched.
This is not a symmetry but a change of variables that brings the theory into a frame that looks like general relativity.
The passage from the Jordan frame to the Einstein frame is of this type.
Since the literature does not use these terms consistently, keep this distinction in mind.

\section{Weyl Geometry}
\label{sec:WG}

\subsection{Weyl connection}

Let $\tilde V^\mu(x+dx)$ denote the vector obtained by parallel transporting a vector $V^\mu(x)$ from the point $x$ to the point $x+dx$, where parallel transport refers to transport determined by the affine connection.
In contrast to Riemannian geometry, the magnitude of the vector is not preserved under parallel transport in Weyl geometry, as described below.
We define
\begin{equation}
    \tilde V^\mu(x+dx)
    \equiv V^\mu(x)-\tilde\Gamma^\mu{}_{\rho\sigma}(x)V^\rho(x)dx^\sigma,
    \label{eq:vector_palla}
\end{equation}
where $\tilde\Gamma^\mu{}_{\rho\sigma}(x)V^\rho(x)dx^\sigma$ is the deviation of $\tilde V^\mu(x+dx)$ from $V^\mu(x)$ caused by parallel transport, and it grows in proportion to the magnitudes of $V^\rho(x)$ and $dx^\sigma$.
The proportionality coefficient $\tilde\Gamma^\mu{}_{\rho\sigma}(x)$ is the affine connection.

A geometry in which the length gauge at each point of spacetime, the Weyl gauge, can be chosen arbitrarily and independently at every point is called Weyl geometry, and the space to which this geometry applies is called a Weyl space.
In a Weyl space, the Weyl gauge generally differs from point to point, so that the magnitude of a vector appears to change under parallel transport.
For a vector $V^\mu(x)$, the square of its magnitude is $\{V(x)\}^2=g_{\mu\nu}(x)V^\mu(x) V^\nu(x)$.
Suppose that the Weyl gauge at each point of the Weyl space has been fixed in advance.
When a vector whose length at the point $x$ is $\{V(x)\}^2$ is parallel transported to the point $x+dx$, we define its length measured in the Weyl gauge at $x+dx$ by
\begin{equation}
    \{\tilde V(x+dx)\}^2
    \equiv\{1-\alpha A_\mu(x)dx^\mu\}\{V(x)\}^2.
    \label{eq:weyl_vector}
\end{equation}
In other words, when the lengths of the vector at $x$ and at $x+dx$ are compared, the length equal to $\{V(x)\}^2$ is $\{\tilde V(x+dx)\}^2$\,, which is $\{V(x)\}^2$ multiplied by $\{1-\alpha A_\mu(x)dx^\mu\}$.
Here, the vector $A_\mu(x)$ represents the rate of dilation of length accompanying the transport and is called the Weyl gauge field, and $\alpha$ is an arbitrary constant.

From Eqs.~\eqref{eq:vector_palla} and \eqref{eq:weyl_vector}, we obtain
\begin{equation}
    \partial_\lambda g_{\mu\nu}+\alpha A_\lambda g_{\mu\nu}
    -g_{\mu\alpha}\tilde\Gamma^\alpha{}_{\nu\lambda}-g_{\alpha\nu}\tilde\Gamma^\alpha{}_{\mu\lambda}
    =0.
    \label{eq:metric_palla}
\end{equation}
Hereafter, we assume that the lower indices of the affine connection are symmetric,
\begin{equation}
    \tilde\Gamma^\rho{}_{\mu\nu}(x)
    =\tilde\Gamma^\rho{}_{\nu\mu}(x).
    \label{eq:torsion_free}
\end{equation}
Then, the affine connection of the Weyl space (hereafter called the Weyl connection) is found to be
\begin{equation}
    \tilde\Gamma^\rho{}_{\mu\nu}
    =\Gamma^\rho{}_{\mu\nu}+C^\rho{}_{\mu\nu},
    \label{eq:weyl_conn}
\end{equation}
where
\begin{equation}
    \Gamma^\rho{}_{\mu\nu}
    =\frac{1}{2}g^{\rho\sigma}(\partial_\mu g_{\nu\sigma}+\partial_\nu g_{\mu\sigma}-\partial_\sigma g_{\mu\nu})
\end{equation}
are the Christoffel symbols, and
\begin{equation}
    C^\rho{}_{\mu\nu}
    =\frac{1}{2}\alpha(\delta^\rho{}_\mu A_\nu+\delta^\rho{}_\nu A_\mu-g_{\mu\nu}A^\rho).
\end{equation}

\subsection{Weyl transformation}

We define the transformation of the length gauge, the Weyl gauge, at each point by
\begin{equation}
    g_{\mu\nu}(x)
    \to\mathrm e^{\xi(x)}g_{\mu\nu}(x),
    \label{eq:weyl_gauge}
\end{equation}
where neither the coordinates $x$ nor the components $V^\mu(x)$ of a vector are changed.

When a quantity $Q(x)$ transforms under the transformation \eqref{eq:weyl_gauge} as
\begin{equation}
    Q(x)
    \to\mathrm e^{n\xi(x)}Q(x),
    \label{eq:weyl_weight}
\end{equation}
we call $Q(x)$ a Weyl covariant quantity of weight $n$.
For instance, $g_{\mu\nu}$ has the weight $1$, $g^{\mu\nu}$ has the weight $-1$, and $\sqrt{-g}$ has the weight $2$.
The components of a vector $V^\mu(x)$ are Weyl invariant, but its magnitude transforms such that $\{V(x)\}^2$ has the weight $1$.
In addition, the Weyl gauge field $A_\mu(x)$ transforms under \eqref{eq:weyl_gauge} as
\begin{equation}
    A_\mu(x)
    \to A_\mu(x)-\frac{1}{\alpha}\partial_\mu\xi(x).
    \label{eq:weyl_trans_A}
\end{equation}
The set of transformations \eqref{eq:weyl_gauge},~\eqref{eq:weyl_weight}, and \eqref{eq:weyl_trans_A} is called a Weyl transformation.
The Weyl connection~\eqref{eq:weyl_conn} is Weyl invariant.

\subsection{Weyl covariant derivative}

For a type $(\gamma\,,\omega)$ tensor, which is a Weyl covariant quantity of weight $n$, we define the Weyl covariant derivative as
\begin{equation}
\begin{aligned}
    \tilde\nabla_\lambda T^{\alpha\beta\cdots\gamma}{}_{\mu\nu\cdots\omega}
    &\equiv(\partial_\lambda+n\alpha A_\lambda)T^{\alpha\beta\cdots\gamma}{}_{\mu\nu\cdots\omega}\\
    &\quad+\tilde\Gamma^\alpha{}_{\lambda\rho}T^{\rho\beta\cdots\gamma}{}_{\mu\nu\cdots\omega}
    +\cdots
    +\tilde\Gamma^\gamma{}_{\lambda\rho}T^{\alpha\beta\cdots\rho}{}_{\mu\nu\cdots\omega}\\
    &\quad-\tilde\Gamma^\rho{}_{\lambda\mu}T^{\alpha\beta\cdots\gamma}{}_{\rho\nu\cdots\omega}
    -\cdots
    -\tilde\Gamma^\rho{}_{\lambda\omega}T^{\alpha\beta\cdots\gamma}{}_{\mu\nu\cdots\rho}.
    \label{eq:Weyl_derivative}
\end{aligned}
\end{equation}
$\tilde\Gamma^\alpha{}_{\mu\nu}$ is the connection associated with general coordinate transformations, while $A_\mu$ is the connection associated with the Weyl transformations.
Using Eq.~\eqref{eq:metric_palla}, the Weyl covariant derivative of the metric tensor becomes
\begin{equation}
    \tilde\nabla_\lambda g_{\mu\nu}
    =0,
    \quad
    \tilde\nabla_\lambda g^{\mu\nu}
    =0.
\end{equation}
Thus, Weyl geometry is metric-compatible with respect to the Weyl covariant derivative that includes the weight term $n\alpha A_\lambda$.
In contrast, the covariant derivative built solely from the Weyl connection is non-metric.
From Eq.~\eqref{eq:metric_palla}, one has $\nabla^{(\tilde\Gamma)}_\lambda g_{\mu\nu}=-\alpha A_\lambda g_{\mu\nu}$\,.
This shows that the length of a vector changes under parallel transport.

We also define the covariant derivative of a type $(\gamma\,,\omega)$ tensor with respect to the Christoffel symbols, the Weyl covariant derivative at $\alpha=0$, by
\begin{equation}
\begin{aligned}
    \nabla_\lambda T^{\alpha\beta\cdots\gamma}{}_{\mu\nu\cdots\omega}
    &\equiv\partial_\lambda T^{\alpha\beta\cdots\gamma}{}_{\mu\nu\cdots\omega}\\
    &\quad+\Gamma^\alpha{}_{\rho\lambda}T^{\rho\beta\cdots\gamma}{}_{\mu\nu\cdots\omega}
    +\cdots
    +\Gamma^\gamma{}_{\rho\lambda}T^{\alpha\beta\cdots\rho}{}_{\mu\nu\cdots\omega}\\
    &\quad-\Gamma^\rho{}_{\mu\lambda}T^{\alpha\beta\cdots\gamma}{}_{\rho\nu\cdots\omega}
    -\cdots
    -\Gamma^\rho{}_{\omega\lambda}T^{\alpha\beta\cdots\gamma}{}_{\mu\nu\cdots\rho}
    \label{eq:LV_derivative}
\end{aligned}
\end{equation}
for which $\nabla_\lambda g_{\mu\nu}=0$.

\subsection{Geometrical quantities in Weyl geometry}

We define the Weyl gauge curvature by
\begin{equation}
    F_{\mu\nu}
    \equiv\partial_\mu A_\nu-\partial_\nu A_\mu,
    \label{eq:Weyl_gauge_curvature}
\end{equation}
which is invariant under the transformation \eqref{eq:weyl_trans_A}.
It satisfies antisymmetry $F_{\mu\nu}=-F_{\nu\mu}$ and the identity
\begin{equation}
    \partial_\lambda F_{\mu\nu}+\partial_\mu F_{\nu\lambda}+\partial_\nu F_{\lambda\mu}=0.
\end{equation}

The $(1,3)$ curvature tensor of Weyl geometry, corresponding to the Riemann curvature tensor in Riemannian geometry, is defined as
\begin{equation}
    \tilde R^\alpha{}_{\beta\mu\nu}
    \equiv\partial_\mu\tilde\Gamma^\alpha{}_{\nu\beta}
    -\partial_\nu\tilde\Gamma^\alpha{}_{\mu\beta}
    +\tilde\Gamma^\alpha{}_{\mu\rho}\tilde\Gamma^\rho{}_{\nu\beta}
    -\tilde\Gamma^\alpha{}_{\nu\rho}\tilde\Gamma^\rho{}_{\mu\beta}.
    \label{eq:Weylian_Riemann_tensor}
\end{equation}
Using the covariant derivative with respect to the Christoffel symbols in Eq.~\eqref{eq:LV_derivative}, it can also be written as
\begin{equation}
    \tilde R^\alpha{}_{\beta\mu\nu}
    =R^\alpha{}_{\beta\mu\nu}
    +\nabla_\mu C^\alpha{}_{\nu\beta}-\nabla_\nu C^\alpha{}_{\mu\beta}
    +C^\alpha{}_{\mu\rho}C^\rho{}_{\nu\beta}-C^\alpha{}_{\nu\rho}C^\rho{}_{\mu\beta},
\end{equation}
where $R^\alpha{}_{\beta\mu\nu}$ is the Riemann tensor.
The tensor $\tilde R^\alpha{}_{\beta\mu\nu}$ does not possess the same index symmetries as the Riemann tensor.
Indeed,
\begin{equation}
\begin{aligned}
    \tilde R_{\beta\alpha\mu\nu}
    &=-\tilde R_{\alpha\beta\mu\nu}+\alpha g_{\alpha\beta}F_{\mu\nu},\\
    \tilde R_{\alpha\beta\nu\mu}
    &=-\tilde R_{\alpha\beta\mu\nu},\\
    \tilde R_{\mu\nu\alpha\beta}
    &=\tilde R_{\alpha\beta\mu\nu}
    -\frac{\alpha}{2}
    (g_{\alpha\beta}F_{\mu\nu}-g_{\mu\nu}F_{\alpha\beta}+g_{\nu\alpha}F_{\mu\beta}
    -g_{\nu\beta}F_{\mu\alpha}+g_{\mu\beta}F_{\nu\alpha}-g_{\mu\alpha}F_{\nu\beta}).
\end{aligned}
\end{equation}
It satisfies the identities
\begin{equation}
    \tilde R^\alpha{}_{\beta\mu\nu}+\tilde R^\alpha{}_{\mu\nu\beta}+\tilde R^\alpha{}_{\nu\beta\mu}
    =0
\end{equation}
and
\begin{equation}
    \tilde\nabla_\lambda\tilde R^\alpha{}_{\beta\mu\nu}
    +\tilde\nabla_\mu\tilde R^\alpha{}_{\beta\nu\lambda}
    +\tilde\nabla_\nu\tilde R^\alpha{}_{\beta\lambda\mu}
    =0.
\end{equation}
Contracting the first and third indices in Eq.~\eqref{eq:Weylian_Riemann_tensor}, the $(0,2)$ curvature tensor of Weyl geometry, corresponding to the Ricci tensor in Riemannian geometry, is obtained as
\begin{equation}
    \tilde R_{\mu\nu}
    =R_{\mu\nu}
    +\frac{1}{2}\alpha\left(F_{\mu\nu}-2\nabla_\nu A_\mu-g_{\mu\nu}\nabla_\alpha A^\alpha\right)
    -\frac{1}{2}\alpha^2\left(g_{\mu\nu}A_\alpha A^\alpha-A_\mu A_\nu\right)\,,
\end{equation}
where $R_{\mu\nu}$ is the Ricci tensor and its symmetry property reads
\begin{equation}
    \tilde R_{\nu\mu}
    =\tilde R_{\mu\nu}-2\alpha F_{\mu\nu}.
\end{equation}
The scalar curvature in Weyl geometry, corresponding to the Ricci scalar in Riemannian geometry, is
\begin{equation}
    \tilde R
    =R-3\alpha\nabla_\alpha A^\alpha-\frac{3}{2}\alpha^2A_\alpha A^\alpha,
    \label{eq:weyl_scalar}
\end{equation}
where $R$ is the Ricci scalar.

\section{Higher-curvature gravity in Weyl geometry}
\label{sec:HCG}

\subsection{Jordan frame}

We consider an action of the form
\begin{equation}
    S
    =\int\!\mathrm d^4x\,\sqrt{-g}\,\mathcal L.
\end{equation}
For the action to be Weyl invariant, $\mathcal L$ must be a Weyl covariant quantity of weight $-2$, because $\sqrt{-g}$ has the weight $2$.
Since $\tilde R$ has the weight $-1$, a general function of $\tilde R$ cannot be used to construct a Weyl-invariant action.
Among functions of $\tilde R$ alone, the only exception is $\tilde R^2$, which has the weight $-2$ by itself.
We therefore introduce a scalar field $\phi$ as a Weyl covariant quantity of weight $-1/2$.
With the help of this scalar field, one can build quantities of weight $-2$, such as $\phi^2\tilde R$.
Likewise,
\begin{equation}
    \mathcal R
    \equiv\frac{\tilde R}{\phi^2}
\end{equation}
has the weight $0$, i.e., it is Weyl invariant, and since $\phi^4$ has the weight $-2$, the combination $\phi^4f(\mathcal R)$ has the weight $-2$ for an arbitrary function $f$.
Therefore, the action of Weyl-invariant $f(\mathcal R)$ gravity is
\begin{equation}
    S
    =\int\!\mathrm d^4x\,\sqrt{-g}\,
    \left[
    \phi^4f(\mathcal R)
    -\frac{1}{4}F_{\mu\nu}F^{\mu\nu}
    -\frac{\zeta}{2}g^{\mu\nu}\tilde\nabla_\mu\phi \tilde\nabla_\nu\phi
    \right],
    \label{eq:action_f(R)}
\end{equation}
where $\zeta$ is a dimensionless parameter.
Since $\phi$ has the weight $-1/2$, $\tilde\nabla_\mu\phi=(\partial_\mu-\alpha A_\mu/2)\phi$.
We keep $\zeta$ explicitly because a redefinition of the scalar field can absorb its magnitude but not its sign, provided that the functional form of $f$ is left arbitrary, as also emphasized in Ref.~\cite{Tang:2020ovf}.
In other words, the sign of $\zeta$ changes the physical content of the theory.
For the vector field, by contrast, the magnitude of the coefficient of $F_{\mu\nu}F^{\mu\nu}$ can be absorbed into $\alpha$, and its sign is fixed by requiring a healthy kinetic term.
The case $\zeta=0$, in which $\phi$ has no kinetic term, is also allowed and includes Weyl's original quadratic theory.
Note also that $\phi$ cannot be brought to $\phi=0$ by a Weyl transformation with finite $\xi$, and that the scalar field always appears as $\phi^2$, so that of the two branches $\phi>0$ and $\phi<0$ we can choose $\phi>0$ without loss of generality.

Using an auxiliary field $\chi$, let us rewrite the action \eqref{eq:action_f(R)} as
\begin{equation}
    S
    =\int\!\mathrm d^4x\,\sqrt{-g}\,
    \left[
    \phi^4f(\chi)+\phi^4 f'(\chi)(\mathcal R-\chi)
    -\frac{1}{4}F_{\mu\nu}F^{\mu\nu}
    -\frac{\zeta}{2}g^{\mu\nu}\tilde\nabla_\mu\phi \tilde\nabla_\nu\phi
    \right].
    \label{eq:action_chi}
\end{equation}
Varying the action with respect to $\chi$ gives
\begin{equation}
    \phi^4 f''(\chi)(\mathcal R-\chi)
    =0,
\end{equation}
where a prime $'$ denotes differentiation with respect to the argument.
Assuming $f''(\chi)\neq0$ we obtain $\chi=\mathcal R$, so that the actions \eqref{eq:action_f(R)} and \eqref{eq:action_chi} are equivalent.

Let us define new variables
\begin{equation}
    \Phi
    \equiv\phi^2f'(\chi),
    \quad
    u(\chi)\equiv
    \chi f'(\chi)-f(\chi),
\end{equation}
where $\Phi$ has the weight $-1$ and $u(\chi)$ is Weyl invariant.
In terms of these quantities, the action \eqref{eq:action_chi} becomes
\begin{equation}
    S
    =\int\!\mathrm d^4x\,\sqrt{-g}\,
    \left[
    \Phi\tilde R
    -\phi^4 u(\chi)
    -\frac{1}{4}F_{\mu\nu}F^{\mu\nu}
    -\frac{\zeta}{2}g^{\mu\nu}\tilde\nabla_\mu\phi \tilde\nabla_\nu\phi
    \right].
\end{equation}
Writing this action in terms of quantities of Riemannian geometry by inserting Eq.~\eqref{eq:weyl_scalar} and integrating by parts, we obtain
\begin{equation}
    S
    =\int\!\mathrm d^4x\,\sqrt{-g}\,
    \left[
    \Phi R-\phi^4 u(\chi)
    -\frac{1}{4}F_{\mu\nu}F^{\mu\nu}
    -\frac{\zeta}{2}g^{\mu\nu}\nabla_\mu\phi \nabla_\nu\phi
    +\alpha A^\alpha\nabla_\alpha\psi
    -\frac{\alpha^2}{2}A_\alpha A^\alpha\psi
    \right],
    \label{eq:action_psi}
\end{equation}
where
\begin{equation}
    \psi
    \equiv\frac{\zeta}{4}\phi^2+3\Phi
    =\phi^2\Psi(\chi),
    \quad
    \Psi(\chi)
    \equiv\frac{\zeta}{4}+3 f'(\chi)\,.
    \label{eq:psi}
\end{equation}
It is crucial that the same combination $\psi$ appears both in the term linear in $A_\mu$ and in the term quadratic in $A_\mu$.
Because $\Psi(\chi)$ is Weyl invariant, the Weyl weight of $\psi$ is carried entirely by $\phi^2$.
Therefore, we can redefine the vector field $A_\mu$ as
\begin{equation}
    B_\mu
    \equiv A_\mu-\frac{1}{\alpha\psi}\nabla_\mu\psi
    \label{eq:def_B}
\end{equation}
Since $\psi$ has the weight $-1$, under a Weyl transformation $\psi^{-1}\nabla_\mu\psi\to\psi^{-1}\nabla_\mu\psi-\nabla_\mu\xi$, so that $B_\mu\to B_\mu$.
Thus, $B_\mu$ is a Weyl-invariant massive vector field.
The Weyl gauge curvature~\eqref{eq:Weyl_gauge_curvature} is unchanged:
\begin{equation}
    F_{\mu\nu}
    =\partial_\mu A_\nu-\partial_\nu A_\mu
    =\partial_\mu B_\nu-\partial_\nu B_\mu.
\end{equation}
The field $\psi$ plays the role of the Stueckelberg field.
We emphasize that no gauge fixing was performed.
Written in terms of $B_\mu$, the action \eqref{eq:action_psi} reads
\begin{equation}
    S
    =\int\!\mathrm d^4x\,\sqrt{-g}\,
    \left[
    \Phi R
    -\phi^4 u(\chi)
    -\frac{1}{4}F_{\mu\nu}F^{\mu\nu}
    -\frac{1}{2}\alpha^2\psi B_\mu B^\mu
    -\frac{\zeta}{2}\nabla_\mu\phi \nabla^\mu\phi
    +\frac{1}{2\psi}\nabla_\mu\psi\nabla^\mu\psi
    \right].
    \label{eq:action_B}
\end{equation}

We now discuss the no-ghost condition for the graviton.
From the action \eqref{eq:action_B}, the coefficient of the gravitational term $R$, namely $\Phi=\phi^2 f'(\chi)$, must be positive, that is, $\Phi>0$.
Because $\phi^2>0$, this requires
\begin{equation}
    f'(\chi)>0.
    \label{eq:c1}
\end{equation}
This corresponds to the no-ghost condition $F'(R)>0$ for $F(R)$ gravity in Riemannian geometry.

\subsection{Einstein frame}

In what follows, we assume $f'(\chi)>0$.
We define the conformal transformation
\begin{equation}
    \hat g_{\mu\nu}
    \equiv\Omega^2(x)g_{\mu\nu},
    \quad
    \Omega^2(x)
    \equiv\frac{2\Phi}{M_p^2}>0.
    \label{eq:conformal_trans}
\end{equation}
This is a redefinition of the metric field.
Unlike a Weyl transformation \eqref{eq:weyl_gauge},~\eqref{eq:weyl_weight}, and \eqref{eq:weyl_trans_A}, no other field is transformed.
We have chosen $\Omega^2$ such that $\hat g_{\mu\nu}$ is Weyl invariant and the coefficient of the gravitational term is $M_p^2/2$.
Indeed, $g_{\mu\nu}$ has the weight $1$ and $\Phi=\phi^2 f'(\chi)$ has the weight $-1$, so that $\Omega^2$ has the weight $-1$ and $\hat g^{\mu\nu}$ is Weyl invariant.
We then have
\begin{equation}
    \sqrt{-\hat g}
    =\Omega^4\sqrt{-g},
    \quad
    \hat g^{\mu\nu}
    =\Omega^{-2}g^{\mu\nu},
    \quad
    \hat R
    =\Omega^{-2}(R-6\Omega^{-1}\nabla_\mu\nabla^\mu\Omega).
\end{equation}
Then, the gravitational term $\sqrt{-g}\,\Phi R$ transforms to
\begin{equation}
    \sqrt{-g}\,\Phi R
    =\frac{M_p^2}{2}\sqrt{-\hat g}\,
    \left(\hat R
    -6\hat\nabla_\mu\ln\phi\hat\nabla^\mu\ln\phi
    -6\hat\nabla_\mu\ln\phi\hat\nabla^\mu\ln f'
    -\frac{3}{2}\hat\nabla_\mu\ln f'\hat\nabla^\mu\ln f'
    \right).
\end{equation}
Using $\psi=\phi^2\Psi$, $2\Psi-\zeta/2=6f'$ and $\nabla_\mu\ln\Psi=(3f'/\Psi)\nabla_\mu\ln f'$, the kinetic terms of the scalar sector can be expressed in terms of logarithmic derivatives as
\begin{equation}
\begin{aligned}
    \mathcal L_k
    &=-\frac{\zeta}{2}\nabla_\mu\phi \nabla^\mu\phi+\frac{1}{2\psi}\nabla_\mu\psi\nabla^\mu\psi\\
    &=\phi^2\left(
    6f'\nabla_\mu\ln\phi\nabla^\mu\ln\phi
    +6f'\nabla_\mu\ln\phi\nabla^\mu\ln f'
    +\frac{9f'^2}{2\Psi}\nabla_\mu\ln f'\nabla^\mu\ln f'
    \right),
\end{aligned}
\end{equation}
and thus, $ \sqrt{-g}\,\mathcal L_k$ transforms as
\begin{equation}
    \sqrt{-g}\,\mathcal L_k
    =\frac{M_p^2}{2}\sqrt{-\hat g}
    \left(6\hat\nabla_\mu\ln\phi\hat\nabla^\mu\ln\phi
    +6\hat\nabla_\mu\ln\phi\hat\nabla^\mu\ln f'
    +\frac{9f'}{2\Psi}\hat\nabla_\mu\ln f'\hat\nabla^\mu\ln f'\right).
\end{equation}
The vector field is unaffected, $B_\mu=\hat B_\mu$, and its kinetic term is conformally invariant, so that $\sqrt{-g}F_{\mu\nu}F^{\mu\nu}=\sqrt{-\hat g}\hat F_{\mu\nu}\hat F^{\mu\nu}$.
Performing the conformal transformation of the action \eqref{eq:action_B}, the terms proportional to $\hat\nabla_\mu\ln\phi\hat\nabla^\mu\ln\phi$ and to $\hat\nabla_\mu\ln\phi \hat\nabla^\mu\ln f'$ cancel identically against the contribution of the gravitational term, and we obtain
\begin{equation}
    \hat S
    =\int\!\mathrm d^4x\,\sqrt{-\hat g}\,
    \left[
    \frac{M_p^2}{2}\hat R
    -\frac{1}{4}\hat F_{\mu\nu}\hat F^{\mu\nu}
    -\frac{1}{2}m_B^2(\chi)\hat B_\mu \hat B^\mu
    -\frac{3\zeta M_p^2}{16}\frac{f''^2}{\Psi f'^2}\hat\nabla_\mu\chi\hat\nabla^\mu\chi
    -\hat V(\chi)
    \right],
    \label{eq:action_Einstein}
\end{equation}
where
\begin{equation}
    m_B^2(\chi)
    \equiv\frac{M_p^2\alpha^2\Psi(\chi)}{2f'(\chi)},
    \quad
    \hat V(\chi)
    \equiv\frac{M_p^4}{4}\frac{u(\chi)}{f'^2(\chi)}
    =\frac{M_p^4}{4}\frac{\chi f'-f}{f'^2}.
\end{equation}
Both are Weyl invariant and functions of $\chi$ only.
Using $u'=\chi f''$, we also have
\begin{equation}
    \frac{\mathrm d\hat V(\chi)}{\mathrm d\chi}
    =\frac{M_p^4}{4}\frac{u'f'-2uf''}{f'^3}
    =\frac{M_p^4}{4}\frac{f''(2f-\chi f')}{f'^3}.
\end{equation}
An important consequence of the action \eqref{eq:action_Einstein} is that no kinetic term for $\phi$ survives.
This is a consequence of Weyl invariance.
All the fields $\hat g_{\mu\nu}$, $\hat B_\mu$, and $\chi$ appearing in the action \eqref{eq:action_Einstein} are Weyl invariant; therefore, the action cannot depend on the compensator $\phi$ at all.
However, the degrees of freedom of $\phi$ have not disappeared.
Through the Stueckelberg field $\psi=\phi^2\Psi$, they have been absorbed into $B_\mu$ by Eq.~\eqref{eq:def_B}.
Therefore, the propagating degrees of freedom are the massless graviton, the massive vector $B_\mu$ and, for $\zeta\neq0$, the scalar $\chi$, i.e., the scalaron.

Given $f'(\chi)>0$, the condition for the absence of a tachyon in the vector sector is $\Psi>0$.
The condition for the absence of a ghost in the scalar sector is $\zeta/\Psi>0$.
The two conditions are simultaneously satisfied if and only if
\begin{equation}
    \zeta>0.
    \label{eq:c2}
\end{equation}
Indeed, if $\zeta>0$ then $\Psi=\zeta/4+3f'$ is a sum of positive quantities and hence $\Psi>0$.
If, on the other hand, $\zeta<0$, then the sign of the scalar kinetic term requires $\Psi<0$, but then $m_B^2<0$.
The scalar condition alone would allow the branch $\zeta<0$, $\Psi<0$, i.e. $0<f'<-\zeta/12$, which corresponds to the solutions with $\zeta<0$ considered in Ref.~\cite{Tang:2020ovf}; however, this branch is excluded by the no-tachyon condition of the vector.
Once $\zeta>0$ is imposed, $\Psi>0$ holds automatically, so that the vector sector imposes no condition beyond Eq.~\eqref{eq:c2}.
Moreover, the case $m_B=0$, i.e. $\Psi=0$, is automatically excluded by the conditions \eqref{eq:c1} and \eqref{eq:c2}.

The boundary case $\zeta=0$ deserves a separate comment.
The kinetic term of $\chi$ in the action \eqref{eq:action_Einstein} then vanishes identically, and $\chi$ becomes an auxiliary field determined algebraically by its equation of motion.
No ghost appears, the scalaron does not propagate, and the spectrum consists of the graviton and the massive vector only, with $m_B^2=3\alpha^2M_p^2/2>0$ since $\Psi=3f'$.
For $f=b\mathcal R^2$, this is Weyl's original quadratic action, whose broken phase is the Einstein--Proca theory with this vector mass~\cite{Ghilencea:2018dqd,Ghilencea:2019jux}.
The applications in Refs.~\cite{Yang:2022icz,Burikham:2023bil,Haghani:2023nrm,Harko:2024fnt,Oancea:2023ylb} are based on this action without a kinetic term for the scalar, so their spectrum contains no propagating scalaron.
Therefore, condition \eqref{eq:c2} is the condition for a healthy propagating scalaron, and $\zeta\geq0$ is the condition for the absence of ghosts in the scalar sector and tachyons in the vector sector.

\subsection{Canonical scalar}
In what follows, we assume $f'(\chi)>0$ and $\zeta>0$.
Introducing the new variable $w\equiv\ln f'$ and the new field $\hat\sigma$ defined by
\begin{equation}
    \frac{\mathrm d\hat\sigma}{\mathrm dw}
    =M_p\sqrt{\frac{3\zeta}{8\Psi(w)}},
    \quad
    \Psi(w)
    =\frac{\zeta}{4}+3\mathrm e^w,
    \quad
    \kappa
    =\frac{M_p}{2}\frac{\mathrm dw}{\mathrm d\hat\sigma}
    =\sqrt{\frac{2\Psi}{3\zeta}},
\end{equation}
we obtain
\begin{equation}
    \hat\sigma
    =M_p\sqrt{\frac{3}{2}}\ln\frac{2\sqrt{\Psi}-\sqrt{\zeta}}{2\sqrt{\Psi}+\sqrt{\zeta}}.
    \label{eq:sigma}
\end{equation}
The integration constant was chosen such that $\hat\sigma\to0$ as $\Psi\to\infty$, that is, as $f'(\mathcal R)\to\infty$.
Note that $f'(\mathcal R)>0$ and $\zeta>0$ imply $\Psi>\zeta/4$.
Thus, the argument of the logarithm lies between $0$ and $1$, and the range of $\hat\sigma$ is $-\infty<\hat\sigma<0$.
Equation~\eqref{eq:sigma} can be inverted.
Setting $\theta=\hat\sigma/(\sqrt{6}M_p)$, we obtain
\begin{equation}
    \Psi(\hat\sigma)
    =\frac{\zeta}{4}\coth^2\theta,
    \quad
    f'(\hat\sigma)
    =\frac{\zeta}{12\sinh^2\theta},
    \quad
    \kappa(\hat\sigma)
    =\frac{|\coth\theta|}{\sqrt{6}}.
\end{equation}
In the limit $\theta\to-\infty$ one has $\Psi\to\zeta/4$ and $f'\to0$, whereas in the limit $\theta\to0^{-}$ one has $\Psi\to\infty$ and $f'\to\infty$.

In terms of the canonical scalar field $\hat\sigma$, the action \eqref{eq:action_Einstein} becomes
\begin{equation}
    S
    =\int\!\mathrm d^4x\,\sqrt{-\hat g}\,
    \left[
    \frac{M_p^2}{2}\hat R
    -\frac{1}{4}\hat F_{\mu\nu}\hat F^{\mu\nu}
    -\frac{1}{2}m_B^2\hat B_\mu \hat B^\mu
    -\frac{1}{2}\hat\nabla_\mu\hat\sigma\hat\nabla^\mu\hat\sigma
    -\hat V(\hat\sigma)
    \right],
    \label{eq:action_4}
\end{equation}
where
\begin{equation}
    m_B^2(\hat\sigma)
    =\frac{3}{2}\alpha^2M_p^2\cosh^2\left(\frac{\hat\sigma}{\sqrt{6}M_p}\right),
    \quad
    \hat V(\hat\sigma)
    =\frac{M_p^4}{4}\frac{\mathcal R(\hat\sigma)f'(\hat\sigma)-f(\mathcal R(\hat\sigma))}{f'^2(\hat\sigma)}.
    \label{eq:vector_mass_sigma}
\end{equation}
Here, $\mathcal R(\hat\sigma)$ is obtained by solving $f'(\mathcal R)=\zeta/(12\sinh^2\theta)$ for $\mathcal R$.
All the physical content of the theory is contained in $m_B^2(\hat\sigma)$ and $\hat V(\hat\sigma)$.
Note that $m_B^2(\hat\sigma)$ contains neither $\zeta$ nor $f'$, and it is manifestly positive.
Since $\cosh^2\theta\geq1$, the vector mass is moreover bounded from below, $m_B^2\geq3\alpha^2M_p^2/2$, independently of $f$ and $\zeta$, and the bound is approached as $\hat\sigma\to0^-$.

As for $\hat V(\hat\sigma)$, we have
\begin{equation}
    \frac{\mathrm d\hat V(\hat\sigma)}{\mathrm d\hat\sigma}
    =\frac{M_p^3\kappa}{2}\frac{2f-\mathcal Rf'}{f'^2}.
    \label{eq:dV}
\end{equation}
Since $\kappa>0$, the stationary point, $\mathrm d\hat V(\hat\sigma)/\mathrm d\hat\sigma=0$, is determined by
\begin{equation}
    \mathcal Rf'(\mathcal R)
    =2f(\mathcal R),
    \label{eq:stationarity_condition}
\end{equation}
which corresponds to the de Sitter condition $RF'(R)=2F(R)$ of $F(R)$ gravity in Riemannian geometry.
Let $\mathcal R_0$ be the value satisfying the stationarity condition and $\hat\sigma_0$ the corresponding value of $\hat\sigma$.
Differentiating Eq.~\eqref{eq:dV} once more and evaluating the result at the stationary point, we obtain
\begin{equation}
    m_{\hat\sigma}^2
    =\left.\frac{\mathrm d^2\hat V}{\mathrm d\hat\sigma^2}\right|_{\hat\sigma_0}
    =M_p^2\kappa^2\left[\frac{1}{f''(\mathcal R_0)}-\frac{\mathcal R_0}{f'(\mathcal R_0)}\right].
\end{equation}
Since $\kappa^2=2\Psi/(3\zeta)>0$, the absence of a tachyon in the scalar sector is equivalent to
\begin{equation}
    \frac{1}{f''(\mathcal R_0)}>\frac{\mathcal R_0}{f'(\mathcal R_0)}.
    \label{eq:absence_tachyon}
\end{equation}
This inequality cannot be rearranged without knowing the sign of $f''$.
The sign of $\mathcal R_0$, which is determined by the nature of the background, is also required.

Since $f(\mathcal R_0)=\mathcal R_0f'(\mathcal R_0)/2$ at the stationary point from Eq.~\eqref{eq:stationarity_condition}, the value of the potential reduces to
\begin{equation}
    \hat V(\mathcal R_0)
    =\frac{M_p^4}{8}\frac{\mathcal R_0}{f'(\mathcal R_0)}.
    \label{eq:V_R0}
\end{equation}
This value acts as an effective cosmological constant, $ \hat V(\mathcal R_0)=M_p^2\Lambda$, and since $f'(\mathcal R_0)>0$ the sign of $\Lambda$ is the sign of $\mathcal R_0$.
There are three possibilities: de Sitter spacetime for $\mathcal R_0>0$, Minkowski spacetime for $\mathcal R_0=0$, and anti-de Sitter spacetime for $\mathcal R_0<0$.
The Minkowski case can be characterized more explicitly.
$\hat V(\mathcal R_0)=0$ requires $f(\mathcal R_0)=0$, which together with $\mathcal R_0f'(\mathcal R_0)=2f(\mathcal R_0)$ and $f'(\mathcal R_0)\neq0$ gives $\mathcal R_0=0$ and $f(0)=0$ as a necessary and sufficient condition.

Consider first a de Sitter background, $\mathcal R_0>0$.
If $f''(\mathcal R_0)$ were negative, the left-hand side of Eq.~\eqref{eq:absence_tachyon} would be negative, while the right-hand side would be positive, which is impossible. 
Hence,
\begin{equation}
    f''(\mathcal R_0)>0.
    \label{eq:c3}
\end{equation}
Multiplying Eq.~\eqref{eq:absence_tachyon} by $f''(\mathcal R_0)>0$ then gives $\mathcal R_0f''/f'<1$, while $\mathcal R_0>0$ and Eq.~\eqref{eq:c3} give $\mathcal R_0f''/f'>0$.
The two bounds combine into the single inequality
\begin{equation}
    0<\frac{\mathcal R_0f''(\mathcal R_0)}{f'(\mathcal R_0)}<1,
    \label{eq:dS_condition}
\end{equation}
which is the same structure that appears in the vacuum of Riemannian $F(R)$ gravity.
We remark that in Riemannian $F(R)$ gravity, the condition $F''>0$ is often quoted as an independent requirement, its rationale being the effective mass of the scalaron in a background containing matter, that is, the Dolgov--Kawasaki instability.

The remaining two cases are treated in the same manner.
On the Minkowski background, $\mathcal R_0=0$, the right-hand side of Eq.~\eqref{eq:absence_tachyon} vanishes, and the condition reduces to
\begin{equation}
    f''(0)>0,
    \label{eq:Min_condition}
\end{equation}
with $m_{\hat\sigma}^2=M_p^2\kappa^2/f''(0)$.
On the anti-de Sitter background, $\mathcal R_0<0$, Eq.~\eqref{eq:absence_tachyon} is automatically satisfied for $f''(\mathcal R_0)>0$, whereas for $f''(\mathcal R_0)<0$ it demands $\mathcal R_0f''/f'>1$, so that Eq.~\eqref{eq:absence_tachyon} should be used in its original form.

Regarding $m_{\hat\sigma}^2$, we stress that it has been evaluated for the canonically normalized scalar field $\hat\sigma$ in the Einstein frame, and at a stationary point of the potential, $\mathrm d\hat V/\mathrm d\hat\sigma=0$.
When one discusses the effective mass away from a stationary point, care must be taken because $\hat\sigma$ is a nonlinear function of $\mathcal R$.

To summarize, for Weyl geometric  $f(\mathcal R)$ gravity, the no-ghost and no-tachyon conditions for an arbitrary $f$ are \eqref{eq:c1},~\eqref{eq:c2},~\eqref{eq:absence_tachyon}, the last of which takes the form~\eqref{eq:dS_condition} on the de Sitter background and the form~\eqref{eq:Min_condition} on the Minkowski background.
Conditions \eqref{eq:c1} and \eqref{eq:absence_tachyon} have the same algebraic structure as in Riemannian $F(R)$ gravity,
whereas \eqref{eq:c2} is specific to Weyl geometric $f(\mathcal R)$ gravity.
The latter is a consequence of the fact that the scalar field $\phi$ is necessarily introduced once Weyl invariance is imposed for an arbitrary $f$.
Once \eqref{eq:c2} is imposed, the Weyl vector field imposes no further condition.
Note that the no-tachyon condition of the vector is essential here: the no-ghost condition of the scalar alone would admit $\zeta<0$.

We emphasize that this coincidence is algebraic and that the two theories are not equivalent.
Weyl geometric $f(\mathcal R)$ gravity contains, in addition to the graviton and the scalaron, a massive vector field whose mass \eqref{eq:vector_mass_sigma} depends on the value of the scalaron.
It contains a free parameter $\zeta$ which has no counterpart in the Riemannian theory, and in the variable $\hat\sigma$ its scalar field space ends at $\hat\sigma=0$, at a finite distance, corresponding to $f'\to\infty$, that is, to $\phi\to0$ at fixed $\Phi$.

\section{Quadratic curvature gravity}
\label{sec:QCG}

As an example, we consider
\begin{equation}
    f(\mathcal R)
    =\lambda+a\mathcal R+b\mathcal R^2,
\end{equation}
where $\lambda$, $a$, and $b$ are constants.
Under the field replacement $\phi\to k\phi$ with a constant $k$ in the action, the action \eqref{eq:action_f(R)} keeps its form with $(\lambda, a, b, \zeta)\to(k^4\lambda, k^2a, b, k^2\zeta)$, so that for $a\neq0$ only the combinations $\lambda/a^2, b$ and $\zeta/a$ are physical.
The magnitude of $a$ can be absorbed into $\phi$, but its sign cannot.
In this case, $f'=a+2b\mathcal R$, $f''=2b$, $u=b\mathcal R^2-\lambda$ and $\Psi=\zeta/4+3(a+2b\mathcal R)$.
For $a\neq0$, the stationarity condition \eqref{eq:stationarity_condition} gives $\mathcal R_0=-2\lambda/a$ and $f'(\mathcal R_0)=(a^2-4b\lambda)/a$, and hence
\begin{equation}
    \Lambda
    =\frac{\hat V(\mathcal R_0)}{M_p^2}
    =-\frac{M_p^2}{4}\frac{\lambda}{a^2-4b\lambda}.
\end{equation}
Since $f'(\mathcal R_0)>0$, Eq.~\eqref{eq:V_R0} shows that the sign of $\Lambda$ is that of $\mathcal R_0=-2\lambda/a$:
$\lambda/a<0$ gives de Sitter, $\lambda=0$ gives Minkowski, and $\lambda/a>0$ gives anti-de Sitter.

Since $1/f''(\mathcal R_0)-\mathcal R_0/f'(\mathcal R_0)=a^2/[2b(a^2-4b\lambda)]$, the no-tachyon condition~\eqref{eq:absence_tachyon} reads $b(a^2-4b\lambda)>0$, while the no-ghost condition~\eqref{eq:c1} at the stationary point reads $(a^2-4b\lambda)/a>0$.
Their product gives $ab>0$.
Collecting these results, the conditions~\eqref{eq:c1},~\eqref{eq:c2}, and~\eqref{eq:absence_tachyon} reduce to
\begin{equation}
    \zeta>0,\quad ab>0,\quad b(a^2-4b\lambda)>0,
    \label{eq:hcg_condition}
\end{equation}
together with $a+2b\mathcal R>0$ over the range of $\mathcal R$ under consideration.
There are two branches.
For $a>0$ and $b>0$, Eq.~\eqref{eq:hcg_condition} requires $\lambda<a^2/(4b)$, and all three backgrounds are allowed.
On the de Sitter background, $\mathcal R_0f''/f'=-4b\lambda/(a^2-4b\lambda)$, so that the upper bound of Eq.~\eqref{eq:dS_condition} is satisfied identically and the lower bound reduces to $b\lambda<0$, that is, $\lambda<0$.
On the Minkowski background, Eq.~\eqref{eq:Min_condition} reads $2b>0$, and on the anti-de Sitter background Eq.~\eqref{eq:absence_tachyon} holds automatically for $0<\lambda<a^2/(4b)$.
For $a<0$ and $b<0$, Eq.~\eqref{eq:hcg_condition} requires $\lambda<a^2/(4b)<0$, so that $\mathcal R_0<0$ and only the anti-de Sitter background is allowed, on which $f''<0$ and Eq.~\eqref{eq:absence_tachyon} holds in the form $\mathcal R_0f''/f'>1$.
Therefore, accelerated expansion selects the branch $a>0$ and $b>0$ with $\lambda<0$, and for $a=1$ the conditions reduce to $b>0$, $\zeta>0$ and $1-4b\lambda>0$.

Using $\mathcal R=(f'-a)/(2b)$, we find
\begin{equation}
    \hat V(\hat\sigma)
    =\frac{M_p^4}{4\zeta^2}\left[\frac{(\zeta-12a\sinh^2\theta)^2}{4b}-144\lambda\sinh^4\theta\right].
\end{equation}
In the limit $\lambda\to0$, it reduces to $M_p^4/(16b)[1-(12a/\zeta)\sinh^2\theta]^2$.

At the stationary point $\mathcal R_0$, the masses of the vector and scalar are
\begin{equation}
    \frac{m_B^2}{M_p^2}
    =\frac{\zeta+12f'_0}{8f'_0}\alpha^2,
    \quad
    \frac{m_{\hat\sigma}^2}{M_p^2}
    =\frac{a(\zeta+12f'_0)}{12b\zeta f'_0},
    \label{eq:mass_hcg}
\end{equation}
where $f'_0\equiv f'(\mathcal R_0)=(a^2-4b\lambda)/a$ and their mass ratio is $m_B^2/m_{\hat\sigma}^2=3\zeta\alpha^2b/(2a)$.
The positivity of $m_{\hat\sigma}^2$ is equivalent to $ab>0$, in accordance with Eq.~\eqref{eq:hcg_condition}.
When  $|b\lambda|\ll a^2$ and $\zeta/a$ is fixed, $m_B^2$ is controlled by $\alpha$ and $m_{\hat\sigma}^2$ by $b$, and inside the stability region, both states are necessarily present, the scalaron owing its propagation to $\zeta>0$.

The linear term in $f$ is essential for the scalaron mass, since $m_{\hat\sigma}^2$ in Eq.~\eqref{eq:mass_hcg} is proportional to $a$.
For $a=0$, the stationarity condition~\eqref{eq:stationarity_condition} requires $\lambda=0$.
For the purely quadratic function $f=b\mathcal R^2$ with $\zeta>0$, one has $f'=2b\mathcal R$ and $u=b\mathcal R^2$, so that the potential in Eq.~\eqref{eq:vector_mass_sigma} is constant, $\hat V=M_p^4/(16b)$.
The stationarity condition~\eqref{eq:stationarity_condition} is then satisfied identically.
Because the potential is constant, every value of $\hat\sigma$ is a stationary point, and the second derivative of the potential vanishes, so that $m_{\hat\sigma}^2=0$.
The scalaron propagates but is massless, and the vacuum is de Sitter with $\Lambda=M_p^2/(16b)$ for $b>0$.
Thus, the propagation of the scalaron requires $\zeta>0$, whereas its mass requires the linear term.
The vector mass \eqref{eq:vector_mass_sigma}, on the other hand, depends on $\hat\sigma$ even when the potential is flat.

\section{Conclusion}
\label{sec:concl}

In this study, we constructed $f(\mathcal R)$ gravity based on Weyl geometry.
Since $\sqrt{-g}$ is a Weyl covariant quantity of weight $2$, the Lagrangian density $\mathcal L$ must be of weight $-2$.
Because $\tilde R$ is of weight $-1$, a general function of $\tilde R$ does not yield a Weyl-invariant action, and one is forced to introduce a scalar field $\phi$ of weight $-1/2$.
Using this $\phi$, one can build the Weyl-invariant quantity $\mathcal R=\tilde R/\phi^2$, and then $\phi^4f(\mathcal R)$ provides a Weyl-invariant action for an arbitrary $f$.

For Weyl geometric $f(\mathcal R)$ gravity, we have established the stability conditions for arbitrary $f$.
They are $f'>0$, $\zeta>0$ (with $\zeta=0$ a degenerate case without a propagating scalaron), and Eq.~\eqref{eq:absence_tachyon}, which on the de Sitter background takes the form $f''(\mathcal R_0)>0$ and $0<\mathcal R_0f''(\mathcal R_0)/f'(\mathcal R_0)<1$.
These follow, respectively, from the gravitational term in the Jordan frame, from the signs of the scalar kinetic term and of the vector mass term in the Einstein frame, and from the second derivative of the potential of the canonical scalar at its stationary point.
Because $\Psi=\zeta/4+3f'$ enters the conditions only through its sign and is a sum of positive terms once $\zeta>0$, the Weyl vector field imposes no further condition.
For the same reason, the case $\Psi=0$, in which the Weyl vector would be massless, is never realized.
In particular, the branch $\zeta<0$ is excluded because the Weyl vector is tachyonic there.
In the boundary case $\zeta=0$, the kinetic term of $\chi$ vanishes in the Einstein frame, so that $\chi$ is an auxiliary field and the scalaron does not propagate.
Then, the dynamical degrees of freedom consist only of the graviton and the massive vector field with $m_B^2=3\alpha^2M_p^2/2$.
This case includes Weyl's original quadratic theory.
Therefore, a propagating scalaron requires both $f''\neq0$ and $\zeta>0$.
In the quadratic example $f=\lambda+a\mathcal R+b\mathcal R^2$, the conditions reduce to $\zeta>0$, $ab>0$, and $b(a^2-4b\lambda)>0$, and the scalaron mass is proportional to the coefficient $a$ of the linear term.
For the purely quadratic function $f=b\mathcal R^2$ with $\zeta>0$, the potential is flat, and the scalaron is massless.
Thus, the propagation of the scalaron requires $\zeta>0$, whereas its mass requires the linear term.

Along the way, we have shown that the Weyl vector couples to the scalar sector only through the single combination $\psi$ of the scalar fields, so that the Weyl vector, shifted by the logarithmic gradient of $\psi$, becomes the Weyl-invariant massive vector $B_\mu$ of Eq.~\eqref{eq:def_B} without any gauge fixing, and no kinetic scalar-vector mixing remains in the Einstein frame.

We conclude with an outlook.
In this study, we have dealt with the signs of the mass and kinetic terms around maximally symmetric backgrounds.
It will be necessary to study perturbations on an FLRW background.
There is also room to consider the case in which matter fields are coupled.
Finally, it would be interesting to extend the function to a form $f(\tilde R^\mu{}_{\nu\rho\sigma},\phi^2 g_{\mu\nu})$ as a generic higher-curvature gravity based on Weyl geometry.

\bibliography{refs}

\end{document}